
\def\=#1{\bar #1}
\def\~#1{\widetilde #1}
\def\.#1{\dot #1}
\def\^#1{\widehat #1}
\def\"#1{\ddot #1}

\def \pn{\par\noindent}

\def \ep{\epsilon}
\def \be{\beta}
\def \ga{\gamma}
\def \de{\delta}
\def \om{\omega}
\def \Om{\Omega}
\def \a{\alpha}
\def \th{\theta}
\baselineskip=0.8 cm

\def \ii{\int_{-\infty}^{+\infty}}

\centerline
{\bf Modifying the onset of homoclinic chaos.}
\centerline
{\bf  Application to a bistable  potential}
\bigskip\bigskip\bigskip
\centerline {G. Cicogna and L. Fronzoni}
\bigskip\bigskip
\centerline{\it Dipartimento di Fisica dell'Universit\`a - p.a
Torricelli 2 -  56126 PISA, Italy}

\bigskip\bigskip\bigskip

\pn
PACS 05 45
\bigskip\bigskip\bigskip
\pn
{\tt Phys.Rev. E, to appear}
\vskip 3cm
\pn
{\bf Abstract}
\bigskip
\pn
We analyze, by means of Melnikov method, the possibility of
modifying the threshold of homoclinic chaos in general 1-dimensional
problems, by introducing small periodic resonant modulations. We
indicate in particular a prescription in order to increase the
threshold (i.e. to prevent chaos), and consider then its
application to the bistable Duffing-Holmes potential. All results
are confirmed both by numerical and by analog simulations, showing
that small modulations can in fact sensibly influence the onset of
chaos.


\vfill\eject

The problem of "controlling chaos" has received great attention in
recent years [1-4]. Even if, for many authors, this term means
in general the stabilization of unstable orbits, we are
concerned here more specifically with the modification of the
threshold for the onset of chaos. One of the simplest and most
interesting methods used in order to modify (possibly to increase)
the threshold of chaos which appears in the presence of
homoclinic orbits, is  the introduction of a periodic modulation of
the parameters describing the unperturbed potential. This
possibility has been analyzed for both  the Duffing-Holmes and the
Josephson-junction potential [2,3].

In the present note, we want to propose some generalizations of this
idea: we will consider first a generic 1-dimensional problem by
means of Melnikov theory [5-6], and obtain a simple criterion about
the "correct" choice of this modulation; next, we will apply this
indication to the bistable Duffing-Holmes potential, introducing
{\it two} independent modulations: this will allow us to confirm
both the general   results and the possibility of modifying the
onset of chaos. All the theoretical discussion is well supported
both by numerical and by analog simulations.

Let us start by considering the general case of a 1-dimensional
"equation  of motion" for the real variable $x=x(t)$:
$$\"x=f(x)-\de \.x+\ga \cos \om t +\ep g(x) \cos(\Om t+\th)
\eqno(1)$$
where for $\ep=0$ we have the standard problem of a periodically
forced and linearly damped motion, whereas the additional term with
$\ep\ne 0$  takes into account the presence of general modulating
terms.  We assume as usual $f=-dV/dx$,
where the unperturbed potential $V=V(x)$ has a maximum point at
$x=x_0$  and a  homoclinic orbit, that we indicate by
$$q=q(t) \eqno(2)$$
doubly asymptotic for $t\to\pm\infty$ to $x_0$. In order to obtain a
theoretical estimation of the effect of the last term of Eq. (1)
to the threshold of chaos,  let us write down the Melnikov function
$M(t_0)$ for the problem (1): taking into account that
$q(t)=q(-t)$, it is easily seen that $M(t_0)$ acquires the form
$$\eqalign {&- M(t_0)= \cr
& \de\int_{-\infty}^{+\infty}\.q^2(t)\ dt + \ga\sin\om t_0
\ii \.q(t) \sin\om t\ dt\ + \cr
& \ep \sin(\Om t_0+\th)\ii \.q(t)\ g\big(q(t)\big)\sin\Om t\ dt \
\equiv \ \de J_\de+\ga J_\ga \sin \om t_0 + \ep J_\ep \sin(\Om t_0+
\th)} \eqno(3)$$
At the resonant case, i.e. when $\om=\Om$, this can be written
$$-M(t_0)=\de J_\de +\ga K \sin(\om t_0+\a)\eqno(4)$$
with
$$K=\Big|J^2_\ga + \Big({\ep J_\ep\over{\ga}}\Big)^2 + 2
\Big({\ep\over{\ga}}\Big) J_\ga J_\ep \cos\th\Big|^{1/2}\eqno(4')$$
The condition for the appearance of chaos according to Melnikov
criterion, i.e. that  $M(t_0)$ has simple zeros, becomes now (it is
not restrictive to assume $\ga>0$, whereas $\de>0$ for physical
reasons, and clearly $J_\de>0$):
$$\ga K > \de J_\de \eqno(5)$$
With fixed damping $\de$, we can then use the modulation term in
order  to modify  the threshold of chaos: here, this amounts to
modify the range of the forcing amplitudes $\ga$ which do not
produce chaotic responses. Now, according to Eqs. (4-5), the
maximum increasing of this threshold is obtained by choosing
$\th=0$ and the sign of $\ep$ according to the following
prescription:  $$\ep > 0\quad ({\rm resp.} < 0)\quad {\rm if }\quad
J_\ga J_\ep < 0 \quad ({\rm resp.} > 0)\eqno(6)$$
(or, which is the same, $\ep>0$ in any case, $\th=0$ if
$J_\ga J_\ep < 0$ and $\th=\pi$ if $J_\ga J_\ep >0$).

With this choice for $\ep$, and observing that the amplitude $|\ep|$ of the
modulation is usually very small (here we only require
$|\ep|<\ga |J_\ga/J_\ep|$), then the above condition (5) becomes
$$\ga > {\de J_\de\over{|J_\ga|}}+\Big|\ep{J_\ep\over{J_\ga}}
\Big|\eqno(7)$$
or also
$${\ga\over{\de}} > R(\ep)=R_0 + |\ep| {|J_\ep|\over{\de |J_\ga|}}
\eqno(8)$$
where
$$R_0={J_\de\over{|J_\ga|}}\eqno(8')$$
is the "Melnikov ratio" in the case $\ep =0$ of no modulation.

Then Melnikov theory predicts that the presence of modulation
produces, if  the phase of modulation is chosen according to the
above rule, an  increasing  (proportional to the quantity
$|J_\ep/J_\ga|$) of the threshold of chaos.

Before further discussing this result, let us remark incidentally
that  in the case of non-resonant modulations, $\om\ne \Om$, Eq.(3)
suggests that one may expect \big(at most after some time delay of
the order  $\ \sim 1/(\om-\Om)$\big)  a "in-phase" contribution of
both forcing  and modulating terms $\ \sim \ga |J_\ga|+|\ep
J_\ep|$. Then, the above arguments indicate that a lowering of the
threshold, favouring the onset of chaos, is to be expected in this
case. A careful analysis of nonresonant modulations in the case of
Duffing-Holmes potential can be found in ref.[2].
\medskip
We want now to provide a precise test of the above results by
checking  them in the case of Duffing-Holmes potential. We find it
convenient to introduce a small generalization of the above
discussion, by choosing the perturbation $g(x)$ in (1) in the form
of {\it two} independent terms modulating {\it both} the linear
and  the cubic components of the force: this requires the presence
of {\it two}  modulation parameters $\ep_1, \ep_3$, as made in [4],
where a  Duffing-Holmes - type potential was obtained by means of
a  magnetoelastic equipment.  Precisely, we consider the equation
$$\"x=- A x^3(1+\ep_3 \cos\Om_3 t) + Bx(1+\ep_1 \cos \Om_1 t)  -\de
\.x + \ga \cos \om t \eqno(9)$$
where $A,B,\ga,\de >0$, and the signs of $\ep_1, \ep_3$ are for the
moment undetermined. As well known, the unperturbed Duffing-Holmes
potential possesses two homoclinic orbits, given by
$$q^{(\pm)}=\pm\Big({2B\over{A}}\Big)^{1/2} {\rm sech}\ t \sqrt{B}
\eqno(10)$$
where the signs $+$ and $-$ denote respectively the
orbit surrounding  the  minimum of the potential at
$x^{(+)}=+\sqrt{B/A}$ and at $x^{(-)}=-\sqrt{B/A}$. All integrals
appearing in Melnikov function  can be evaluated exactly; in
particular, the integrals  $J^{(+)}_{\ep_1}$
and $J^{(-)}_{\ep_1}$ take the same value when evaluated
respectively along the orbit  $q^{(+)}(t)$ and along $q^{(-)}(t)$,
the same is true for $J^{(+)}_{\ep_3}$ and $J^{(-)}_{\ep_3}$: we
have
$$J^{(\pm)}_{\ep_1}=-2{B^{3/2}\over {A}}\ii\!\! \sinh \ t
\sqrt{B}\  {\rm sech}^3\ t \sqrt{B}\
 \sin\Om_1 t  \ dt= - {\pi B\Om_1^2\over{A}}\ {\rm cosech}
{\pi \Om_1\over{2\sqrt{B}}}  < 0 \eqno(11)$$
$$J^{(\pm)}_{\ep_3}=4{B^{5/2}\over{A}}\ii\!\! \sinh \ t \sqrt{B}\
{\rm sech}^5 \ t \sqrt{B}\ \sin \Om_3 t \ dt= {\pi \Om_3^2\over{6
A}}(\Om_3^2+4 B)\ {\rm cosech} {\pi \Om_3\over{2 \sqrt{B}}}  >
0 \eqno(11')$$
(notice that the results in Eq. (9) of ref.[2] and Eq.
(4) of ref. [4] are  uncorrect).
One has also $J_\de^{(+)}=J_\de^{(-)}$, whereas $J_\ga^{(+)}=
-J_\ga^{(-)}$:
$$J^{(\pm)}_\de={4\over{3}}{B^{3/2}\over{A}} > 0 \ , \qquad\qquad
J_\ga^{(+)}=-\pi\om\sqrt{{2\over{A}}} {\rm sech} {\pi \om\over{2
\sqrt{B}}} < 0 \eqno(12)$$

Thus, Melnikov condition for the appearance of chaos, in the
resonant case $\om= \Om_1=\Om_3$, can be finally written in the
form  $${\ga\over{\de}} > R^{(\pm)}(\ep_1,\ep_3)=R_0\mp\ep_1r_1\pm
\ep_3 r_3\eqno(13)$$ where again $\pm$ distinguish between the two
homoclinic orbits $q^{(\pm)}(t)$,  and
$$R_0={2\sqrt {2} B^{3/2}\over{3 \sqrt{A}\pi\om}}\cosh {\pi
\om\over{2 \sqrt{B}}}\eqno(14)$$
is the well known ratio for the unperturbed Duffing-Holmes
potential [6], and
$$r_1={\om B\over{\de \sqrt{2 A}}}{\rm cotgh} {\pi \om\over{2
\sqrt{B}}} \ , \qquad r_3={ \om (\om^2+4 B)\over{6\de \sqrt{2 A}}}
{\rm cotgh}  {\pi\om\over{2 \sqrt{B}}} \eqno(14')$$

Therefore, we can conclude:
If the motion occurs near the homoclinic orbit $q^{(+)}(t)$,
the best choice in  order to prevent chaos (by increasing the
threshold) is
$$\ep_1 < 0 \qquad {\rm and} \qquad \ep_3 > 0 \eqno
(15)$$
This choice however favours the onset of chaos when the
motion is in the potential well around $x^{(-)}=-\sqrt{B/A}$. The
opposite choice $\ep_1>0,\ \ep_3<0$ would produce exactly opposite
effects. All these results agree with our above discussion (cf. the
signs of $J_\ga^{(\pm)}$ and $J^{(\pm)}_{\ep_1},
J^{(\pm)}_{\ep_3}$).

Numerically, with e.g. $A=B=\om=1, \ \de=0.25$, we obtain
$$r_1 = 3.08 \ , \qquad\qquad r_3 = 2.57\eqno(16)$$
The relatively large numerical values of these coefficients
(compared with $R_0=0.753$) show that the role of modulation in the
Duffing-Holmes potential is  really important: according to
Eqs.(13,16), one may in fact expect that very small $\ep_1, \ep_3$
may considerably influence the onset of chaos. Another interesting
remark is that the introduction of the modulation (clearly the
effect is present also choosing one of the two $\ep_1, \ep_3$ equal
to zero) produces a sort of "dynamical asymmetry" between the two
potential wells. Let us emphasize that it had been already remarked
[7] that a small "geometrical" asymmetry in the double-well
potential sensibly modifies the thresholds of chaos in the two
wells. Precisely, considering an asymmetric potential
$$V(x)= {A\over{4}}x^4-{B\over{2}}x^2+\be x \eqno(17)$$
it can be shown that the Melnikov ratios $R^{(\pm)}(\be)$, at the
first order in  the asymmetry parameter $\be$, are given by
$$R^{(\pm)}(\be) = R_0 \pm \be \rho \eqno(18)$$
where
$$\rho=R_0{\sqrt{2 A}\over{B^{3/2}}}\Big({3\pi\over {4}}-{\om
\over{\sqrt{B}}} {\rm cotgh}{\pi\om\over{2
\sqrt{B}}}\Big)\eqno(19)$$
With $A=B=\om=1$ as before, one has
$\rho=1.35$, showing that the two effects are in fact comparable.
\medskip
We have tested the above theoretical results for the
Duffing-Holmes potential by means of both numerical and analog
simulations. The agreement is globally good  enough. For instance,
choosing $A=B=\om=1$, and  $\de=0.2 ,\ga=0.2$ (i.e. largely within
the chaotic region if no modulation would be present, cf. [6]), it
is sufficient to insert a modulation with $\ep_1=-0.05 ,
\ep_3=+0.04$ in order to obtain a periodic response. Fig. 1 shows
some of these periodic motions, oscillating around $x^{(+)}=1$,
obtained by numerical integration, starting from different initial
conditions.
\smallskip
Analog simulation is another very convenient
and known method (see [9] and Ref.  therein, and [8,10,3]) to
examine nonlinear systems. The analog device we use in this case is
essentially similar  to others already used and discussed elsewhere
(see especially [9,3]). In particular, the modulation is obtained
by means of multipliers operating in the reaction loop, in a
similar way to the case of the modulated Josephson-junction device,
discussed in detail in [3]. The values of the parameters are
obtained by direct measurements on the experimental circuit. In
particular, the damping term $\de$ is deduced from the resonance
band width at the limit of small amplitudes; therefore, this
measure is actually subject to some uncertainties: We obtain
$\de=0.25$ with an estimated error of $\pm 20\%$. The other
parameters (in dimensionless units, as usual) are $B=1;\ 1/\sqrt A
=2.83;\ \om = 1.42\simeq \sqrt{2B}$ (i.e. the frequency of the
small oscillations around the equilibrium points $x^{\pm}$ of the
unperturbed potential). Then, by increasing the forcing amplitude
$\ga$, we look for the threshold values of $\ga$ which produce the
appearance of chaotic responses, for different values of the
amplitudes $\ep_1,\ \ep_3$ of the modulating perturbation. Fig. 2
shows the values we obtain for the ratio $R=\ga/\de$  (where $\ga$
are these threshold values), as a function of $\ep_1$, with
$\ep_3=0$. In agreement with Eq.(13), we obtain a completely
similar behavior with $\ep_1=0$ and varying $\ep_3$. From our
measures, we get the following results
$$R_0=4.3 \ , \qquad r_1=10.1 \ , \qquad r_3=11.2 \eqno (20)$$
to be compared with the theoretical values deduced from Eqs.
(14-14')
$$R_0=2.82 \ ,\qquad r_1=11.63 \ , \qquad r_3=11.66
\eqno (21)$$
The numerical agreement is not perfect; let us
stress, however, that - apart from the uncertainties and
unavoidable errors in the experimental determinations of the
various quantities (see [3] for some short comments on this point)
- in any case one cannot hope that Melnikov method is able to give
a precise determination of the threshold of chaos; rather, a common
and expected result is actually that Melnikov theory indicates  a
somewhat smaller value than the threshold experimentally detected
(see [6,3]). Let us remark on the other hand the better agreement
we find for the values of $r_1, \ r_3$, and in particular the
result $r_1\simeq r_3$ \big(according to Eq.(14'), $r_1=r_3$ for
$\om=\sqrt {2B}$\big), and finally the agreement shown by Fig. 2
with eq.(13). After these remarks, we believe that all the facts
discussed up to now may be considered globally as a rather good
test for both theory and analog experiment.

\bigskip\bigskip\bigskip


\parindent 0pt

{\bf References}

[1] E. Ott, C. Grebogi and J.A. Yorke, Phys. Rev. Lett. {\bf 64},
1196 (1990); A. Hubler, Helv. Phys. Acta {\bf 62}, 343 (1989); T.B.
Fowler, IEEE Trans. Autom. Control {\bf 34}, 201 (1989); B.B. Plapp
and A.W. Hubler, Phys. Rev. Lett. {\bf 65},  2302 (1990); J. Singer
and Y.-Z. Wang, H.H. Bau, Phys. Rev. Lett. {\bf 66}, 1123 (1991);
G. Cicogna, Nuovo Cim. {\bf 105 B}, 813 (1990)

[2] R. Lima and M. Pettini, Phys. Rev. A {\bf 41}, 726 (1990)

[3] G. Cicogna and L. Fronzoni, Phys. Rev. A {\bf 42}, 1901 (1990)

[4] L. Fronzoni, M. Giocondo and M. Pettini, Phys. Rev. A {\bf 43},
6483 (1991)

[5] V. K. Melnikov, Trans. Moscow Math. Soc. {\bf 12}, 1 (1963)

[6] J. Guckenheimer and P.J. Holmes, "Nonlinear Oscillations,
Dynamical Systems, and Bifurcations of Vector Fields" (Springer, New
York, 1983); A.J. Lichtenberg and M.A. Lieberman, "Regular and
Stochastic Motion" (Springer, New York, 1983);  S. Wiggins, "Global
Bifurcations and Chaos" (Springer, New York, 1988)

[7] G. Cicogna and F. Papoff, Europhys. Lett. {\bf 3}, 963 (1987)

[8] D. D'Humieres, M.R. Beasley, B.A. Huberman and A. Libchaber,
Phys. Rev. A {\bf 26}, 3483  (1982); C.A. Hamilton, Rev. Sci.
Instrum. {\bf 43}, 445 (1972); C.K. Bak and N.F. Pedersen, Appl.
Phys. Lett. {\bf 22}, 149 (1973); J.H. Magerlein, Rev. Sci.
Instrum. {\bf 49}, 486 (1978); A. Yagi and I. Kurosawa, Rev. Sci.
Instrum. {\bf 51}, 14 (1980)

[9] F.C. Moon, "Chaotic Vibrations" (J. Wiley, New York, 1987)

[10] L. Fronzoni, in "Noise in Nonlinear Dynamical systems" (ed.
by F. Moss and P.V.E. McClintock) (Cambridge Univ. Press, Cambridge,
1989), vol. 3, p. 222

\bigskip\bigskip\bigskip
\parindent 0pt

{\bf Figure captions}

Fig.1

Numerical solution of Eq.(9) with different initial conditions,
and $A=B=\om=1;\ \de=\ga=0.2$. The modulation terms $\ep_1=-0.05,\
\ep_3=+0.04$ are chosen according to the prescription (15) in order
to remove chaos: indeed, after some short transient, the solution
is periodic.

\bigskip

Fig. 2

Threshold of chaos in the bistable potential [Eq.(9)] as a function
of modulating perturbation. Here, $\ga$ is the experimental value
(via analog simulation) of the threshold; the ratio $R=\ga/\de$ is
plotted vs. the amplitude $\ep_1$ of modulation, with $\ep_3=0$.
The results with $\ep_1=0$ and varying $\ep_3$ are completely
similar, see Eq.(13).

\bye